\documentclass[fleqn,usenatbib,useAMS]{mnras}

\usepackage{graphicx}	% Including figure files
\usepackage{amsmath}	% Advanced maths commands
\usepackage{amssymb}	% Extra maths symbols
\usepackage{multicol}        % Multi-column entries in tables
\usepackage{bm}		% Bold maths symbols, including upright Greek
\usepackage{pdflscape}	% Landscape pages
\usepackage{verbatim}
\usepackage[flushleft]{threeparttable}
\usepackage[dvipsnames]{xcolor}
\usepackage{ulem}
\newcommand{\HH}{\mbox{H$\rm _2$}}

\newcommand{\Dla}{Damped Lyman-$\alpha$}
\newcommand{\lya}{\mbox{${\rm Ly}\alpha$}}

\newcommand{\HI}{H\,{\sc i}}

\newcommand{\CII}{C\,{\sc ii}}

\newcommand{\OI}{O\,{\sc i}}
\newcommand{\SII}{S\,{\sc ii}}
\newcommand{\SiII}{Si\,{\sc ii}}

\definecolor{green}{rgb}{0,0.4,0}

\newcommand{\iap}{Institut d'Astrophysique de Paris, CNRS-SU, UMR\,7095, 98bis bd Arago, 75014 Paris, France}
\newcommand{\ioffe}{Ioffe Institute, {Politekhnicheskaya 26}, 194021 Saint Petersburg, Russia}

\usepackage[T1]{fontenc}
\usepackage{ae,aecompl}

\usepackage{newtxtext,newtxmath}
\graphicspath{{./}{figures/}}
\title[\CII$^{*}$/\CII\ ratio in high-redshift ISM]{\CII$^{*}$/\CII\ ratio in high-redshift DLAs: 
ISM phase separation %, 
drives the observed bimodality of [\CII] cooling rates}

\author[S. A. Balashev et al.]{S. A. Balashev$^{1}$\thanks{Contact e-mail:\href{mailto:s.balashev@gmail.com}{s.balashev@gmail.com}}, K. N. Telikova$^{1}$, P. Noterdaeme$^{2}$ %\newauthor
\\
$^{1}$ \ioffe \\
$^{2}$ \iap \\
}

\date{}

\pubyear{2021}

\begin{document}
\label{firstpage}
\pagerange{\pageref{firstpage}--\pageref{lastpage}}
\maketitle

% Abstract of the paper
\begin{abstract}

We discuss observations of \CII$^*$/\CII\ ratios and cooling rates due to [\CII]~158$\mu$m emission in high-redshift intervening damped Lyman-$\alpha$ systems towards quasars. We show that the observed bimodality in the \CII\ cooling rates actually reflects a bimodality in the \CII*/\CII$-$metallicity plane that can be naturally explained by phase segregation of the neutral medium,  
without invoking differences in star-formation scenarios.
Assuming realistic distributions of the physical parameters to calculate the phase diagrams, we also reproduce qualitatively the metallicity dependence of this bimodality.  
We emphasize that high-z DLAs mostly probe low-metallicity gas ($Z\lesssim 0.1 Z_{\odot}$), where heating is dominated by cosmic rays (and/or turbulence), and not by photoelectric heating. Therefore even if the gas of DLA is predominantly cold (where the cooling is dominated by [\CII]), the excitation of \CII\ can be used to derive the cosmic ray ionization rate (and/or turbulent heating), but not the UV field, as was previously considered. Alternatively, if the gas in DLA is predominantly warm, \CII$^{*}$/\CII\ can be used to constrain its number density. Finally, we also discuss the importance of the ionized medium, which, if also present along the line of sight, can significantly increase the average \CII$^{*}$/\CII\ ratio. 

\end{abstract}

\begin{keywords}
quasars: absorption lines - galaxies: high-redshift - galaxies: ISM
\end{keywords}

%%%%%%%%%%%%%%%%%%%%%%%%%%%%%%%%%%%%%%%%%%%%%%%%%%

%%%%%%%%%%%%%%%%% BODY OF PAPER %%%%%%%%%%%%%%%%%%

% The MNRAS class isn't designed to include a table of contents, but for this document one is useful.
% I therefore have to do some kludging to make it work without masses of blank space.
\begingroup
\let\clearpage\relax
\endgroup
\newpage

%TC:ignore
\section{Introduction}
\label{sec:Introduction}

It is well established that for a wide range of physical conditions, the phase diagram of the neutral interstellar medium consists of two stable phases \citep{Field1969, Wolfire1995}: a cold (with characteristic temperature $T\lesssim100\,K$) and a warm ($T\lesssim10^4\,K$) phase. These phases coexist in pressure equilibrium within some range of thermal pressures, where the denser cold neutral medium (CNM) with small filling factor ($\lesssim 1\%$) is embedded into the more diffuse warm neutral medium (WNM). This two-phase model is a natural consequence of the thermal balance, where the cooling is dominated by line emission (\lya, [\CII]\,158$\mu$m and [\OI]\,63$\mu$m) and heating is primary due to photoelectric effect and cosmic rays \citep{Field1969}. Since the photoelectric heating scales with UV flux, the phase diagram at solar metallicity depends on the star-formation rate \citep{Wolfire2003}. This motivated \citet{Wolfe2003a} to develop a method to constrain the star-formation rates in high-redshift \Dla\ \citep[DLA,][]{Wolfe2005} systems towards quasar (QSO) sightlines, assuming that the photoelectric heating equates the \CII\ cooling rate, itself being estimated through measurements of \CII* column densities \citep[following][]{Pottasch1979}. 
The obtained \CII\ cooling rates present an evident bimodality \citep{Wolfe2008}, which was interpreted by the authors as arising from a bimodality in star-formation regimes in the galaxies associated with the DLAs. 

However, the majority of DLAs probe WNM \citep{Srianand2005, Kanekar2014}, 
which is actually predominantly cooled by Ly$\alpha$ emission \citep[see e.g.][]{Liszt2002, Draine2011} and {not by [\CII] emission, which is the main coolant only in CNM.} 
In particular, studies by \citet{Liszt2002, Wolfe2003a, Srianand2005} and \citet{Neeleman2015} of the \CII*/\CII\ ratio in DLAs mentioned that \CII* is less excited in WNM than in CNM due to difference in densities and temperatures. Even more, \citealt{Srianand2005} have been already shown that since thermal pressures, at which WNM and CNM coexist, is metallicity-dependent, the typical \CII*/\CII\ ratio increase with metallicity decrease.
Emission line studies also indicate that the [\CII]~158$\mu$m emission is indeed typically $\sim20$ times weaker in the WNM than in the CNM \citep[e.g.][]{Pineda2013}. %\SB{I took it from \citep{Pineda2013}, which refer \citep{Wolfire2010}, however, it seems that latter did not provide that number!}
Finally, observations and simulations show that ionized gas can also provide a large fraction (at least in our Galaxy) of \CII\ observed in emission \citep[e.g.][]{Roy2017, RamosPadilla2021}. 

In this letter, we discuss the \CII\ excitation in the low-metallicity gas that is typically associated with DLAs and revisit the origin of the bimodality in the distribution of the observed \CII\ cooling rates. 

\section{\CII\ fine-structure}
\label{sec:CII}

The ground term of singly ionized carbon, \CII\,$\rm (1s^22s^2)2p\,^2P^{\circ}$, splits into two fine-structure levels, $^2P^{\circ}_{1/2}$ (the ground one) and $^2P^{\circ}_{3/2}$ (usually denoted as \CII*). The energy separation between these levels is $63.42$\,cm$^{-1}$ (or $91.25$\,K), and the corresponding fine-structure [\CII]~$\lambda$158$\,\mu$m transition from typically collisionally excited $^2P^{\circ}_{3/2}$ level {\citep{Goldsmith2012}} to the ground state is a key contributor to the cooling of the cold ISM \citep[see e.g.][]{Lagache2018}. Since the collisional excitation coefficient, $C_{12}$, with atomic hydrogen is roughly $\propto T^{0.35}$ (at temperatures $T\gtrsim100$\,K) %\SB{But see Goldsmith+2012 (need to cite) and Barinovs+2005, where they give $\propto T^{0.14}$} \SB{0.14 is only valid in 20-1000 K range, higher 0.35 is more appropriate.}
one can expect that the \CII* collisional excitation ($\propto nC_{12} \propto T^{-0.65} P_{\rm th}$, where $n$ is a number density and $P_{\rm th}$ is a thermal pressure)  will be more efficient in CNM than in WNM, that coexist at similar characteristic thermal pressures. To discuss this difference quantitatively we need properly describe the phase diagram of neutral ISM and its dependence on the external parameters in the medium.

\section{The phase diagram of neutral ISM}
\label{sec:pd}

We calculated the phase diagram for the neutral ISM assuming thermal balance as was done by \citet{Field1969, Wolfire1995, Wolfire2003, Bialy2019}. We considered cooling by Ly$\alpha$ \citep[see e.g.][]{Spitzer1978}, \CII\ and \OI\ fine-structure emission. The latter were calculated by considering excitation of fine-structure levels \citep[see e.g.][]{Klessen2016} using collisional coefficients\footnote{We assume that the main collisional partner in neutral ISM is atomic hydrogen. The presence of molecular hydrogen can slightly change the cooling, but this may become noticeable only for the dense medium, where the molecular fraction is high.} from \citep{Launay1977a, Barinovs2005, Abrahamson2007}. We considered heating by photo electrons, cosmic rays and turbulence dissipation. For the first two mechanisms, we used a standard description \cite[see e.g.][]{Bialy2019} that depends on the electron fraction, obtained using similar calculations as in \citet{Balashev2020}. {The turbulence dissipation heating is not well constrained in diffuse ISM due to its intermittent nature, i.e. variability of energy sources of the turbulence, and lacks direct observations. For simplicity, we assume an averaged $\Gamma_{\rm turb} \sim \gamma_{\rm turb} n$, where $\gamma_{\rm turb}$ has typical values of $\sim10^{-27}$\,erg\,s$^{-1}$ \citep[see e.g.][]{Elmegreen2004, Pan2009}. This gives a comparable heating of the gas as provided by cosmic ray ionization rate (CRIR) of $\rm 10^{-16}\,s^{-1}$.}

\begin{figure}
\includegraphics [trim=0.0cm 0.0cm 0.0cm 0.0cm,width=\columnwidth]{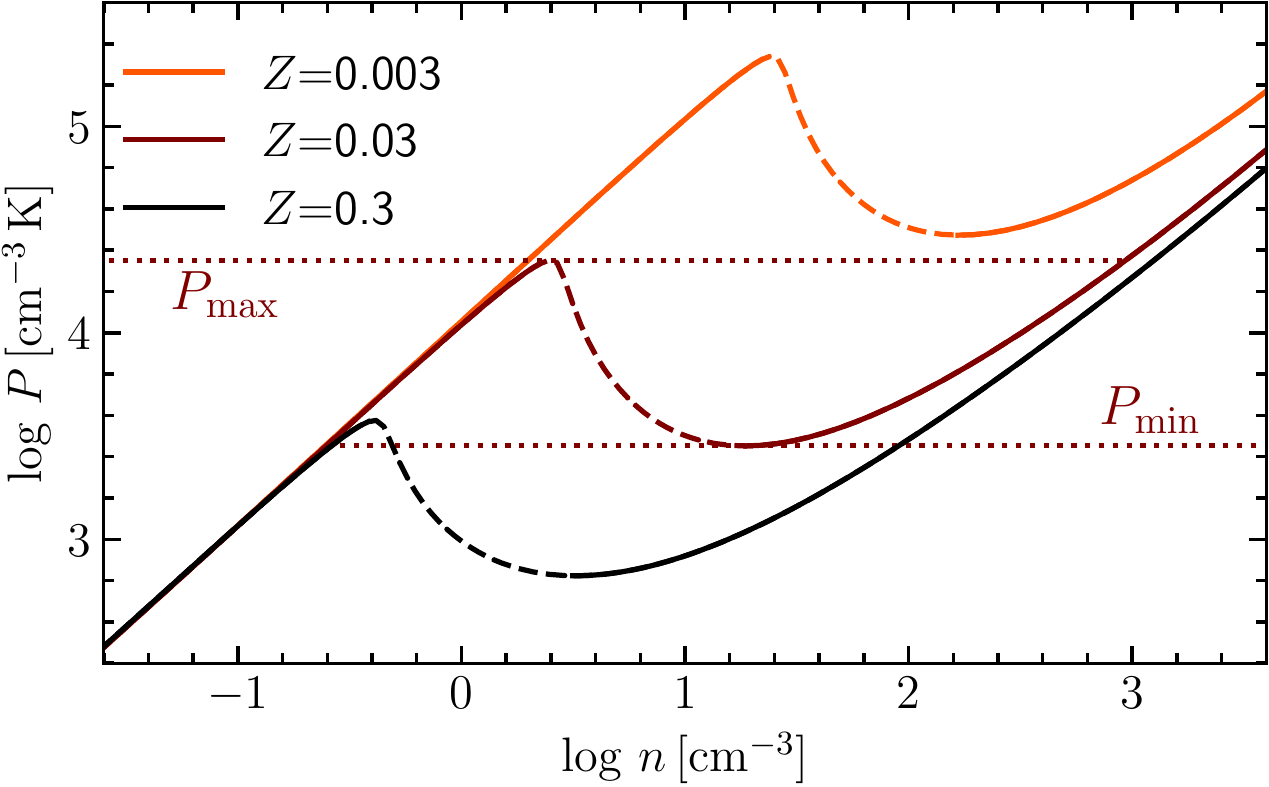}
\caption{The dependence of phase diagram (pressure vs number density) of the neutral ISM on the metallicity. The black, brown and red curves correspond to metallicities $Z=0.3$, $0.03$ and $0.003$, respectively. 
The dashed part of the curves is unstable. The other assumes parameters relevant for the calculation of the phase diagram are given in the text. Dotted horizontal lines indicate the maximal (resp. minimal) pressures where WNM (resp. CNM) can exist, calculated for the {$Z=0.03$} phase diagram. 
} 
\label{fig:pd}
\end{figure}

An important point here is that the phase diagram of the neutral ISM depends on metallicity, as considered in details by several authors \citep{Liszt2002, Bialy2019}\footnote{We did not include the heating and cooling by molecular hydrogen, as considered by \citet{Bialy2019}, since it is not important for the range of metallicity and number density probed by \CII, which is relevant for our study.}. In Fig.~\ref{fig:pd} we show examples of phase diagrams at range of metallicities relevant for our study and calculated for $\rm CRIR=10^{-16}\,s^{-1}$, $\gamma_{\rm turb}=10^{-27}$\,erg\,s$^{-1}$ and UV field $\chi=1$ (in units of Draine field). The dependence on the metallicity is mostly seen as an increase of $P_{\rm min}$ (the minimum pressure for CNM) and $P_{\rm max}$ (the maximal pressure for WNM) with decreasing metallicity. This is due to the fine-structure line cooling that induces the transition between CNM and WNM (and therefore determines $P_{\rm min}$ and $P_{\rm max}$) scaling with metallicity, while the cosmic ray and turbulent heating -- that dominate at low metallicities $Z\lesssim0.1$\footnote{Hereinafter we note metallicity with respect to solar as $Z$.} \citep{Bialy2019} -- do not. The photoelectric heating, which is dominant heating at intermediate metallicities $Z\gtrsim0.1$, is proportional to the dust-to-gas ratio (DTG) and hence scales with metallicity. 
However, observations indicate a non-linear dependence of DTG on $Z$ in particular at low metallicites. %\old{\citep{RemyRuyer2014}, but its exact shape is hard to constrain due to large dispersion}. 
For simplicity, in the following, we consider a single power-law dependence of DTG on $Z$ with a slope $\alpha=2.0$ for the whole range of metallicities \citep[matching the observations from][]{RemyRuyer2014}, and take into account its dispersion (see Section~\ref{sec:results}). 

\section{\CII*/\CII\ ratio}
\label{sec:RC}

\begin{figure}
\includegraphics [width=\columnwidth]{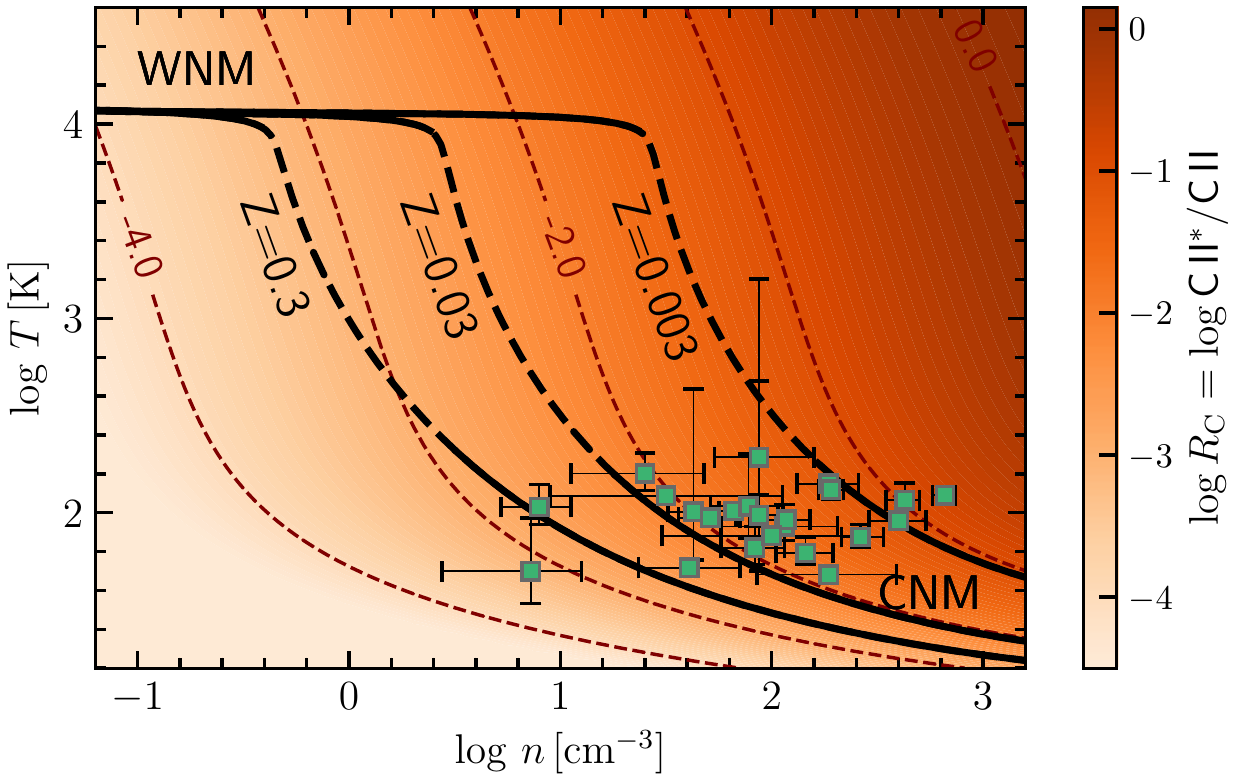}
\caption{\CII*/\CII\ ratios at a function of number density and temperature. The dashed brown curves show the contours of the constant $R_{\rm C}$. The black solid (and dashed) curves correspond to the stable (and unstable) branches of phase diagrams, calculated at the metallicites $Z=0.3$, 0.03, 0.003 (indicated by text labels). Green circles indicate the measurements in H$_2$-bearing components of DLAs (compiled mostly from \citealt{Klimenko2020, Balashev2019} and references therein)
}
\label{fig:nT}
\end{figure}

Under typical ISM conditions, the \CII* level is populated {from the ground level} by collisions {(described by collision rate coefficient $C_{12}$)} and depopulated by spontaneous decay (with the Einstein coefficient $A_{21}=2.29 \times 10^{-6}$\,s$^{-1}$). Hence the population ratio of $^2P^{\circ}_{3/2}$ and $^2P^{\circ}_{1/2}$ levels
\begin{equation}
R_{\rm C} = \frac{n(\text{\CII*})}{n(\text{\CII})} \approx \frac{C_{12}(T) n}{A_{21}} 
\end{equation}
depends on the type of the collision partner (electron, atomic or molecular hydrogen), the number density, $n$, and the temperature, $T$. In Fig.~\ref{fig:nT} we provide calculated $R_{\rm C}$ as a function of $T$ and $n$ {assuming collisions with neutral hydrogen  \citep{Barinovs2005}}\footnote{
Collision with electrons are sub-dominant as long as the  ionization fraction is $\lesssim 10^{-2}$. We discuss the limiting case of ionized medium separately in Section~\ref{sec:IM}.}. 
We also overplot the phase diagrams calculated for different metallicities. One can see that, following the phase diagram, $R_{\rm C}$ remains almost constant in the CNM (since \CII\ fine-structure emission is the main coolant), but it steadily increases with number density (and pressure) in the WNM. Note also that, for a given metallicity, $R_{\rm C}$ remains much higher in the CNM than in the WNM, even at its maximal pressure. 
As metallicity decreases, the typical $R_{\rm C}$ increases for both the CNM and the WNM, owing to the increase of the characteristic number densities and temperatures. Since at $T\gtrsim200$\,K the \CII\ excitation is little sensitive to the temperature (see Fig.~\ref{fig:nT}), it is important to emphasize that 
the observed $R_{\rm C}$ value can be used as a very useful probe of the number density in the WNM and lukewarm thermally unstable gas. In  DLAs, where the temperature can be constrained \citep[][]{Noterdaeme2021b}, \CII\ excitation will additionally provide a measurement of the thermal pressure.

\section{Observations}
\label{sec:DLA}
The dependence of \CII*/\CII\ ratio on the metallicity can be investigated observationally in DLAs using measured column densities. Indeed, in DLAs $N(\text \CII*)$ can be directly constrained from its absorption spectrum (if the line is not significantly saturated), but it is not possible to directly measure the \CII\ column density because of strong saturation of the absorption line. However, one can use the metallicity as a proxy for the abundance of carbon and constrain the \CII*/\CII\ ratio averaged over the velocity components of DLA as
\begin{equation}
R_{\rm C} \approx \frac{N(\text{\CII*})}{N({\text{\HI}}) \left({\rm C}/{\rm H}\right)_\odot \left<Z\right>_{\rm DLA}} \equiv \left<R_{\rm C}\right>_{\rm DLA},
%R_{\rm C} \approx \frac{N(\text{\CII*})}{N({\text{\HI}}) \left({\rm C}/{\rm H}\right)_\odot 10^{\rm [X/H]}} \equiv \left<R_{\rm C}\right>_{\rm DLA},
\label{eq:RC}
\end{equation}
where 
$\left<Z\right>_{\rm DLA}$ is the overall DLA metallicity, averaged over the velocity components\footnote{We neglected depletion for carbon. Although the depletion is about 0.5~dex at solar metallicity, most DLAs in our sample have $\left<Z\right>_{\rm DLA} \lesssim 0.1$, where depletion is much less.} and $\left({\rm C}/{\rm H}\right)_\odot$ is the solar abundance, taken from \citep{Asplund2009}.

\begin{figure*}
\includegraphics [trim=0.5cm 0.5cm 0.5cm 0.5cm,width=1.0\textwidth]{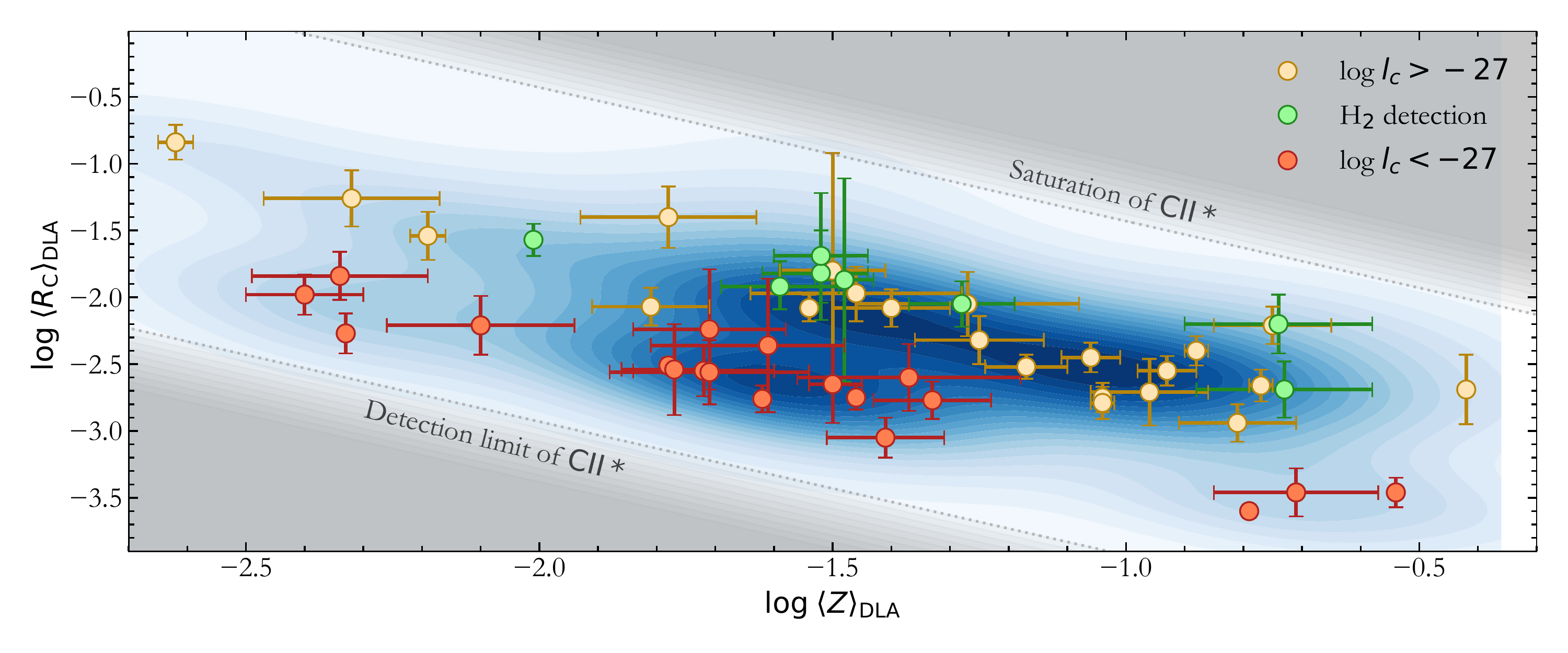}
\caption{
$\left<R_{\rm C}\right>_{\rm DLA}$ as a function of the averaged metallicity observed in DLAs.  
The points with errorbars indicate the observed values towards QSO sightlines. The green points are DLA systems with \HH\ detection. The red points correspond to the systems with low \CII\ cooling rate. The blue contours show the kernel density estimate of the sample. Grey areas correspond to the conditionally defined observational selection of \CII*: the bottom-left region represents the detection limit of \CII*, while the upper-right region corresponds to \CII* saturation of lines, where measurements of $N$(\CII*) are complicated {(see text for details)}. 
}
\label{fig:CII_DLA}
\end{figure*}

In Fig.~\ref{fig:CII_DLA} we plot the measurements of \CII*/\CII\ ratio in DLAs as a function of metallicity\footnote{{The difference in the adopted solar abundances in the literature samples is within the metallicity uncertainties, so we do not take it into account here.}}, using a compilation of data from \citet{Wolfe2008}, \citet{Dutta2014} and \citet{Telikova2021submitted}. We first note that there is a primary selection effect, making some regions of the diagram unavailable either due to weakness or saturation of \CII* lines. {These exclusion regions are illustrated by the greyed areas and limited by the dotted lines 
in Fig.~\ref{fig:CII_DLA}. 
The latter correspond to $\log N$(\CII*) = 12.5 (detection limit) and 15 (saturation limit) represent the low and high end of $N$(\CII*) distribution in the sample, respectively\footnote{{This is only indicative since the exact range of reachable $R_{\rm C}$ for each system depends at least on the spectrum quality and the measured $\log\,N$(\HI) (see equation~\ref{eq:RC}). The latter spans about two dex in the considered sample.}} assuming the median $\log\,N$(\HI)$=21$.
}
 % parameters: $\log\,N$(\HI)$=21$, $\log\,N$(\CII*)$=12.55, 15.63$.
However, one can see that at a given metallicity the \CII*/\CII\ values are generally distributed into two groups of relatively {\it high} and {\it low} \CII*/\CII\ ratios (to highlight this we overplot kernel density estimate of the sample). This separation is evident if we recall the fact that $R_{\rm C} \sim l_{\rm c} / \left<Z\right>_{\rm DLA}$, with evident bimodality in cooling rates, $l_{\rm c}$, distribution \citep[see also ][where the authors discussed metallicity distributions]{Wolfe2008}. 
To illustrate this, we highlight the points with $l_{\rm c} < 10^{-27}\,\rm erg\,s^{-1}$ (considered by \cite{Wolfe2008} as "low"-cooling) by red color. We also differentiate DLAs where molecular hydrogen is detected (green points) and found that these are located among the "high" $R_{\rm C}$ population \citep[see also][]{Srianand2005}. 
Indeed, the presence of molecular hydrogen is a clear indicator of the presence of CNM, which we confirm by measuring the kinetic temperature using the ortho-para ratio of H$_2$, and found $\sim100$\,K for all such DLAs (see Fig.~\ref{fig:nT}).
All this supports our hypothesis that the bimodality in the $R_{\rm C}$-$Z$ distribution originates from phase separation.

\section{Results}
\label{sec:results}

{To further test} whether CNM and WNM drive the bimodality of $R_{\rm C}$-$Z$ distribution, we estimated ranges of $R_{\rm C}$ in WNM and CNM as a function of metallicity using the calculation of the phase diagram, described in Sect.~\ref{sec:pd}. For each metallicity on 
a grid (uniform in log scale) varying from $10^{-3}$ to 1, we generated a distribution of $R_{\rm C}$ in CNM and WNM by sampling the phase diagrams and characteristic pressures. Since DLAs likely probe the overall population of galaxies \citep{Cen2012}, it certainly should have a dispersion in the physical conditions. We assumed lognormal distributions with 0.3 dex dispersion and mean of 1 (in Draine units) 
and $10^{-27}$\,erg\,s$^{-1}$ for the intensity of the UV field and turbulence heating, respectively.  We assumed a CRIR square-law dependence on UV with $\rm CRIR= 10^{-16}$\,s$^{-1}$ corresponding to Draine field \citep[as constrained recently by][and likely due to its association with star-formation]{Kosenko2021}. We also assume the same 0.3 dex dispersion for the DTG along its quadratic dependence on metallicity. 

\begin{figure}
\includegraphics [trim=0.0cm 0.0cm 0.0cm 0.0cm,width=\columnwidth]{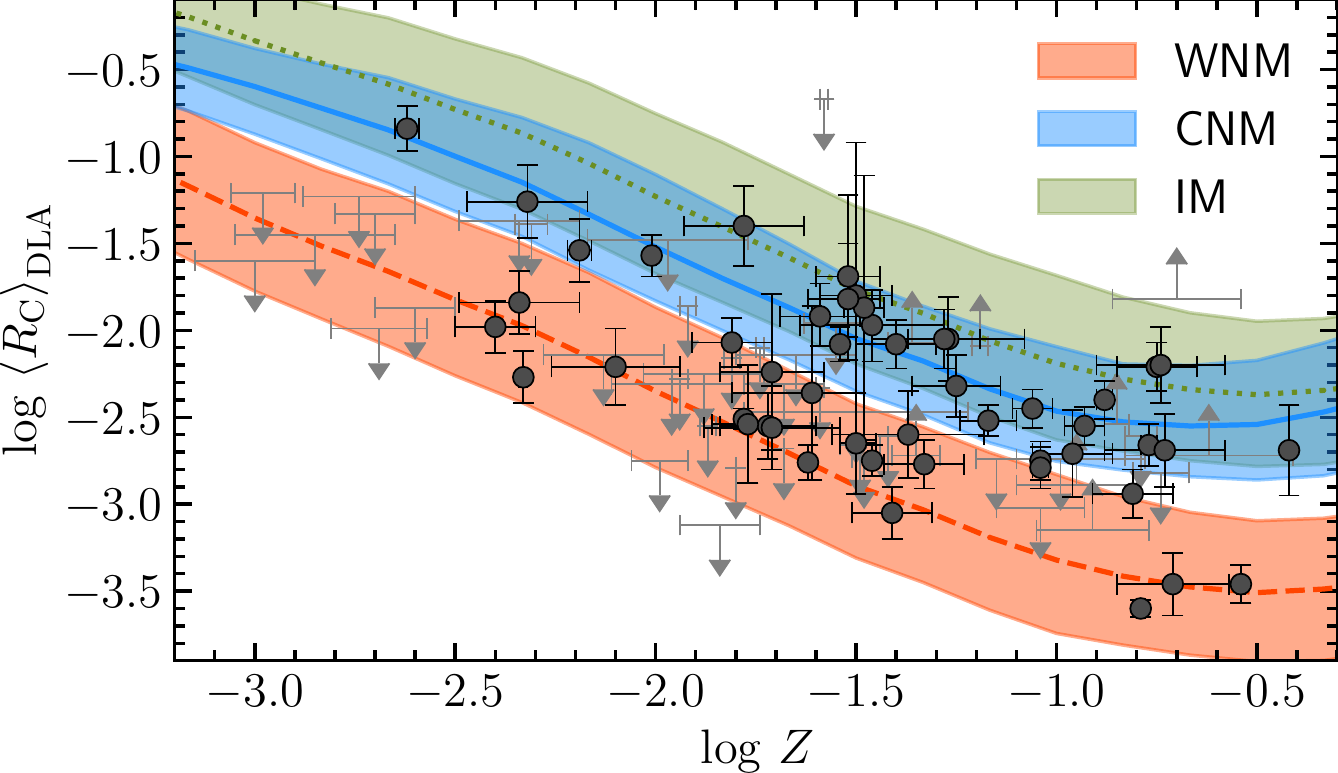}
\caption{\CII*/\CII\ ratio as a function of the metallicity. The black points indicate the observed values towards QSO sightlines (the same as in Fig.~\ref{fig:CII_DLA}{, while the gray markers indicate upper and lower limits \citep[taken from][]{Wolfe2008, Dutta2014}.} The red, blue and {olive} stripes show the ranges where located 68 percent of the sampled distributions calculated for the warm neutral, cold neutral and ionized medium, respectively. The details of the calculation are provided in the text.  
}
\label{fig:CII_Z}
\end{figure}

Then for each phase diagram, we estimated the ranges of $R_{\rm C}$ for CNM and WNM by sampling the pressure uniformly between $P_{\rm min}$ and $P_{\rm max}$ (see Fig.~\ref{fig:pd}) estimated for each phase diagram. In Fig.~\ref{fig:CII_Z} we show the obtained distribution of $R_{\rm C}$  for WNM and CNM as a function of metallicity constructed by such sampling. One can see that the phase separation of CNM and WNM explains $R_{\rm C}$ values measured in DLAs including the general metallicity dependence and the bimodality of $R_{\rm C}$ distribution at each metallicity. We note that the results are quite sensitive to the parameters of the UV, CRIR and turbulence heating distributions, as well as to dependence DTG ratio on the metallicity (especially at the metallicity ranges $Z\gtrsim0.1$) due to obvious dependence of the phase diagram on these parameters. In addition, the results --especially for WNM-- depend on the sampling of pressures (this can be seen in Fig.~\ref{fig:nT}, since $R_{\rm C}$ is very sensitive to $n$ and hence $P$).  
We also note that the observed data points correspond to $R_{\rm C}$ averaged through the neutral gas in the DLA, while we consider only a single phase in our calculations. It is therefore hard to draw robust quantitative conclusions about the physical conditions in DLAs. 

Notwithstanding, we can still qualitatively discuss the sampling of $R_{\rm C}$-$Z$ plane.  
\citet{Wolfe2008} proposed that the difference in the metallicity distribution of the "high-cool" and "low-cool" DLAs originates from different galaxy populations with different SFR regimes. However, as we showed above, the "high-cool" and "low-cool" sub-samples represent DLAs where \CII* column density is predominantly associated with CNM and WNM, respectively. 
Since $R_{\rm C}$ is typically one order of magnitude higher in CNM than in WNM, then even if only a few CNM clouds are intercepted by the line of sight within a given DLA, these CNM clouds will contribute significantly to the total \CII* (and also total \HI) column densities, pushing the overall DLA value close to the pure CNM solution shown in Fig.~\ref{fig:CII_Z}. 
In that sense, the enhanced fraction of high-cool DLAs at $\log Z\gtrsim-1.3$ simply indicates a higher probability for the DLA line of sight to cross CNM gas.
This can be explained by 
the increased number of components at high metallicity \citep[see e.g.][]{Ledoux2006} together with a higher probability for each of them to intersect CNM\footnote{ The minimal pressure required for the existence of CNM decreases with increasing metallicity (see Fig.~\ref{fig:nT}), leading to a decrease of the characteristic number densities and hence an increase of the cross-section of such gas.}. This is confirmed by the higher H$_2$ incidence rate at higher metallicites \citep[][]{Petitjean2006, Noterdaeme2008, Balashev2019}. Interestingly, while H$_2$ always probes CNM, it is detected only in about one forth of the "high" $\left<R_{\rm C}\right>_{\rm DLA}$ (eventhough not all DLAs in the sample have a spectrum that covers the wavelength region where H$_2$ lines are located). 
Selecting high $\left<R_{\rm C}\right>_{\rm DLA}$ could therefore provide an interesting way to uncover a potentially large fraction of CNM that does not produce detectable H$_2$ absorption \citep[as also noted by][]{Krogager2020}.
This fraction is however hard to quantify since our \CII* sample is not fully representative to the whole DLA population.
For example, the typical H$_2$ incidence rate in DLAs is quite low $\lesssim10$\% \citep{Noterdaeme2008, Balashev2018}, while it is found to be higher in this sample ($\gtrsim 15$\%), possibly owing to its skewed \HI\ distribution (the H$_2$ detection rate is known to be much higher at the high end of the \HI\ column density distribution, \citealt{Noterdaeme2015, Balashev2018, Ranjan2020}).

\section{Impact of ionized medium}
\label{sec:IM}

It is important to note that {\it high} \CII*/\CII\ ratios can also arise from ionized medium (IM), which was not taken into account {in our model} before. While studies of our Galaxy \citep[e.g.][]{Lehner2004} as well as modelling \citep[e.g.][]{RamosPadilla2021} all indicate that a large fraction of [\CII]\,158$\mu$m emission can be associated with ionized medium, there are doubts that the same holds for DLAs \citep[e.g. based on the modeling of ion abundances][]{Wolfe2004}. 
However, in addition to neutral gas, the line of sight giving rise to DLAs may also cross ionized gas, owing to in-situ production of ionizing photons in star-forming regions.
Therefore DLAs with total \HI\ column densities sufficiently high for shielding from ionizing photons {in a given individual cloud} can {still} probe ionized phase as well. {Indeed, absorption from metals in high-ionisation stage is generally found associated with DLAs \citep[see e.g.][]{Fox2007}}.
Regarding the ionized phase, the \CII* fine-structure level can be significantly excited by collision with electrons\footnote{The electron excitation typically results in $\sim$ order higher \CII*/\CII\ ratios than the collisions with atomic hydrogen for the same number density.}.
In Fig.~\ref{fig:CII_Z} we plot the expected $R_{\rm C}$ in fully ionized gas assuming a  temperature of $T_{\rm IM}=10^4$\,K and using the same sampling of phase diagram and pressures as for CNM and WNM (see Section~\ref{sec:results}). One can see that if the fraction of IM along the line of sight is relatively high in DLAs (such ionized medium is not detected in \HI), then the integrated \CII*/\CII\ ratio can be much higher than the warm phase limit and mimic the presence of CNM. 
Unfortunately, while we have no constrain on the ionized fraction in DLAs, it is hard to argue how many of the observed \CII*/\CII\ ratios above the warm limit could originate from ionized gas.

\section{Conclusions}

We showed that the previously reported bimodality of observed [\CII] cooling rates ($l_{\rm c}$) in DLAs is essentially the result of combining two effects: (i) the phase separation of the neutral ISM and (ii) the metallicity dependence of this phase splitting. The first results in a much higher $R_{\rm C}={\text \CII*}/{\text \CII}$ ratio in CNM  than in WNM owing to higher collisional excitation. The second maintains this bimodality as a function of $Z$, since  $l_{\rm c} \propto R_{\rm C} \times Z$. Importantly, the metallicity dependence is a consequence from the cooling of CNM being due to the fine-structure line, hence linearly scaling with $Z$, while the heating of the neutral gas becomes insensitive to the metallicity at low-$Z$, where it is dominated by cosmic rays \citep{Bialy2019}.
In the low-metallicity regime, measuring the \CII\ cooling rate (that is the main cooling only if DLA medium is predominantly associated with CNM), permits in fact to constrain the cosmic ray ({and/or} turbulence) heating --assuming thermal balance-- but does not provide {a direct} measurement of the UV field. {We note however that there may be a tight correlation between cosmic rays and UV field \citep{Kosenko2021}}. {If heating is dominated by cosmic rays, then using "high"-\CII* DLAs (likely associated with CNM, where the cooling is due to \CII* emission), we can write $\zeta_{\rm p} = l_{\rm c} E^{-1}_{\rm cr}$, where $E_{\rm cr}\approx 10^{-11}$\,erg in neutral medium \cite[e.g.][]{Bialy2019}. The cosmic rays ionization rates in DLAs are then in the range $\approx 10^{-16}-10^{-15}$\,s$^{-1}$}

While measurements of $R_{\rm C}$ can be used to estimate the {number density in the WNM or lukewarm gas}, 
quantitative usage of the currently available measurements of $R_{\rm C}$ in DLAs is problematic due to non-trivial selection function, and unknown mixing of CNM and WNM along the line of sight. An additional impediment is the possible presence of a large amount of thermally unstable gas \citep[see e.g.][]{Kim2017}, that has $R_{\rm C}$ in between WNM and CNM. Finally, the unknown (and expected to be large) dispersion of the physical parameters (including DTG, CRIR, and turbulence heating) in DLAs also complicates the picture.
Future studies of \CII*/\CII\ ratio should greatly benefit from detailed comparative component-by-component analysis of \CII*\ absorption lines together with those of other species such as \SiII\ or \SII\, since \CII\ is generally not accessible due to saturation of the line. 

Finally, we note that the increase of the \CII* excitation with decreasing metallicity is very important for a proper understanding of the [\CII]~$\lambda$158$\mu$m emissivity function, in particular given the large efforts currently going on to detect galaxies using [\CII] emission at low metallicities \citep[see e.g. dwarf galaxy survey][]{Madden2013} and at high redshifts ($\gtrsim 4$) \citep[e.g.][]{Maiolino2005,
%Iono2006,Cox2011,
Wagg2012}. Furthermore, at these redshifts, one should also take into account the fine-structure excitation by the CMB, which can alter the phase diagram and enhance [\CII]~$\lambda$158$\mu$m line strength from WNM \citep[see e.g.][]{Liszt2002}. 

\section{Data Availability}
The data underlying this article will be shared on reasonable request to the corresponding author.

\section*{Acknowledgements}

We thank the anonymous referee for constructive comments and useful suggestions. This work was supported by RSF grant 18-12-00301 and by the French {\it Agence Nationale de la Recherche} under grant No. 17-CE31-0011-01 ("HIH2").

\bibliographystyle{mnras}
\bibliography{references.bib}
\label{lastpage}
\bsp	% typesetting comment

%TC:endignore
\end{document}